\documentclass[conference]{IEEEtran}
\usepackage{graphicx}
\usepackage{amsmath,amssymb,amsfonts}
\usepackage{algorithmic}
\usepackage{textcomp}
\usepackage[table]{xcolor}
\usepackage{tikz}
\usepackage{array}
\usepackage[hyphens]{url}
\usepackage{hyperref} 
\usepackage{listings}
\usepackage{fancyvrb}
\usepackage{ulem}

\makeatletter
\newcommand\notsotiny{\@setfontsize\notsotiny\@vipt\@viipt}
\makeatother

\IEEEoverridecommandlockouts

\begin{document}
\title{A Continuous Risk Assessment Methodology \\for Cloud Infrastructures}
%
%
\author{
\IEEEauthorblockN{Immanuel Kunz, Angelika Schneider and Christian Banse}
\IEEEauthorblockA{\textit{Fraunhofer AISEC} \\
Garching b. M\"unchen, Germany \\
\{firstname.lastname\}@aisec.fraunhofer.de}
	
}

\maketitle              
\begin{abstract}
Cloud systems are dynamic environments which make it difficult to keep track of security risks that resources are exposed to. Traditionally, risk assessment is conducted for individual assets to evaluate existing threats; their results, however, are quickly outdated in such a dynamic environment. 
%
In this paper, we propose an adaptation of the traditional risk assessment methodology for cloud infrastructures which loosely couples manual, in-depth analyses with continuous, automatic application of their results.
These two parts are linked by a novel threat profile definition that allows to reusably describe configuration weaknesses based on properties that are common across assets and cloud providers. This way, threats can be identified automatically for all resources that exhibit the same properties, including new and modified ones. 
We also present a prototype implementation which automatically evaluates an infrastructure as code template of a cloud system against a set of threat profiles, and we evaluate its performance. 
Our methodology not only enables organizations to reuse their threat analysis results, but also to collaborate on their development, e.g. with the public community. To that end, we propose an initial open-source repository of threat profiles.




\end{abstract}
\section{Introduction}
\label{introduction}


Risk assessment is an essential part of security management, and is commonly used to identify and prioritize risks, and to derive mitigative measures. This process usually identifies individual assets and protection goals, then conducts threat and impact analyses, and finally deduce risk scores. Yet, this sequential risk assessment method assumes a static system for which a once conducted risk assessment can be assumed to be valid for a long period of time. This approach is not suitable for cloud systems that are frequently modified and scaled, e.g. by changing access rights and port configurations, temporarily adding resources like jump servers, and by scaling containers and serverless functions. 


In the past decade, various approaches have been proposed to adapt traditional risk assessment to the cloud. Some automatically assess risks based on known vulnerabilities, e.g. from the National Vulnerability Database~\cite{chih2015adjustable,kamongi2014nemesis}. Others focus on different types of risks, e.g. risk assessment of
compositional cloud systems or various risks beyond security~\cite{djemame2011risk}. 

There are two gaps that remain in the application of risk assessment to the cloud. 
First, expert analyses are still necessary to identify system-specific threats that go beyond known vulnerabilities---yet, their results are quickly outdated. What is required for the cloud is therefore a risk assessment process that adequately integrates both manual, in-depth analyses, and continuous, automated assessments. 

Second, such a process needs to be enabled by a set of technical innovations that allow to automatically discover an up-to-date snapshot of a cloud system and automatically evaluate this snapshot regarding potential threats. Most importantly, however, potential threats need to be described in a highly reusable way that is independent from single resources, and ideally also independent from cloud providers.

In this paper, we redesign the risk assessment process for the cloud by loosely coupling manual expert analyses with the automatic application of the results to cloud resources. For the discovery of existing cloud resources, we use infrastructure as code (IaC) templates, and evaluate them using a policy engine. We furthermore propose threat profiles that can describe sophisticated attack vectors, and which are generically applicable across assets and cloud providers. 
%
When, e.g., a threat is identified in a specific combination of resource configurations, like their location and network configuration, it can then automatically be identified across all such resources---quickly after they have been created or modified. This approach increases the reusability of expert analyses, and allows to share threat descriptions with others across organizations or with the public community. 



The contributions of the paper are the following:
\begin{itemize}
	\item A risk assessment process adapted to continuous cloud assessment, consisting of two modules: An analysis module conducted by experts, and an automated module that continuously applies identified threats to applicable resources,
	\item A novel way to create threat profiles---either based on IaC descriptions or on an ontological description of cloud resources---that make threat descriptions automatically applicable across resources and cloud providers,
	\item A prototype implementation of the automated module: It uses IaC templates and a policy engine to identify applicable threats, 
	\item An initial open-source threat profile repository that cloud users can reuse and contribute to.
\end{itemize}


\section{Related Work}
\label{related_work}

As traditional risk assessment does not work well for the cloud, numerous works have investigated the adaptation of risk assessment for cloud systems.

Note that we define a \textit{cloud provider} as an organization, like Amazon Web Services (AWS) or Microsoft Azure, who provide infrastructure cloud resources, such as virtual machines (VMs). A \textit{cloud user} is an individual or an organization using these resources, e.g. for the purpose of storing data. When a cloud user builds a \textit{cloud service}, e.g. a web shop, on cloud infrastructure, she becomes a \textit{cloud service provider}. 
In contrast, the term \textit{cloud service} is sometimes also used to describe services offered by cloud providers, e.g. AWS S3. 


\subsection{Threat Assessment in the Cloud}
While the topic of \textit{automated} threat analysis of traditional IT-systems has been investigated for a long time, e.g. by Ou et al. \cite{ou2005mulval}, its application to the cloud is more recent. The CloudSafe tool, proposed by An et al.~\cite{an2019cloudsafe}, implements an almost completely automated security analysis of cloud systems. They combine an automatically generated reachability graph of cloud VMs (constructed based on security group definitions) and vulnerabilities of open VM ports acquired from the NIST National Vulnerability Database\footnote{\url{https://nvd.nist.gov}} to evaluate security threats. Other approaches similarly assess risks based on known vulnerabilities~\cite{kamongi2014nemesis}.
Also the approach by Chih and Huang \cite{chih2015adjustable} is based on known vulnerabilities. It is comprised of an offline mode and an online mode where the offline mode is based on historical data about vulnerabilities, whereas the online mode is designed to assess threats at runtime, e.g. based on current Common Vulnerabilities and Exposures (CVEs). 
Note that our approach does not target software vulnerabilities, but rather targets configuration vulnerabilities of a cloud infrastructure.

Note that while we use attack trees for threat modeling in this paper, various other methods for modeling threats exist~\cite{hong2017survey} and can also be used with our approach. 



For the threat analysis in this paper, we propose reusable threat profiles. The concept of reusable security profiles has been applied before as well, e.g. in the Common Criteria approach\cite{ISO15408} where Protection Profiles define threats and other information. It is, however, not targeted at cloud resources.


\subsection{Risk Assessment}
The general approach to risk assessment and management has been covered in various standards~\cite{ISO31000,ISO27005,blank2011guide}. A number of works have also proposed adaptations that are, e.g., more flexible and scalable, or target specific domains~\cite{eichler2015modular,nurse2017security,casola2019toward}. An overview of different risk assessment methods is given by Gritzalis et al.~\cite{gritzalis2018exiting}. 

Concerning risk assessment in the cloud, Djemame et al.~\cite{djemame2011risk} propose a framework that monitors various risks, including security, policy, and legal risks. It also targets different cloud environments, like private, federated, and multi-cloud.
They use a risk database that holds assessments of threats and vulnerabilities for certain assets. Still, their approach is focused on unique assets rather than generically applicable threats. 
%
Their later proposal~\cite{djemame2014risk} features risk assessment both from an infrastructure provider's and a customer's perspective. It overlaps partly with what we propose in this paper, i.e. in their considerations about risk assessment from the customer's perspective during service operation. They do not, however, provide a way to assess risks resource- and vendor-independently, and they do not provide a collaborative way of assessing risks. It is furthermore rather a framework than a concrete proposal of how to discover and evaluate cloud resources' states.

Others base their risk assessment on high-level cloud threats. Saripalli and Walters~\cite{saripalli2010quirc} for example, identify six threat categories for the cloud and calculate their probabilities based on quantitative studies. 
They integrate experts only to evaluate impacts, using the wide-band Delphi method. 

\subsection{Cloud Security Certification}
In our approach we separate expert analyses from the automatic application of their results. In the context of continuous security certification for the cloud, often a similar separation between experts, i.e. auditors, on the one hand, and the automatic verification of results on the other, is employed. 

For instance, Anisetti et al. \cite{anisetti2017semi} propose a framework in which a certification scheme is first defined based on requirements established by a certification authority. Thereafter, the framework verifies gathered evidences of the certification target for consistency with these requirements. 
Krotsiani et al. \cite{krotsiani2015monitoring} develop a monitoring-based approach to continuous certification which also allows for incremental analysis of measurements.

A central challenge in cloud security certification is to evaluate if a set of measurements is sufficient to prove a pre-defined certification requirement.
The goal of our approach is to quickly assess new threats, rather than complying with a pre-defined set of requirements. As such, we do not only decouple expert analyses from the application of their results, but integrate them in periodic re-assessments.

\subsection{Cloud Monitoring Tools}
Commonly used tools for threat modeling and analysis include OWASP Threat Dragon\footnote{\url{https://owasp.org/www-project-threat-dragon}} and Microsoft Threat Modeling Tool\footnote{\url{https://docs.microsoft.com/en-us/azure/security/develop/threat-modeling-tool}}. Using these tools, a data flow diagram (DFD) can be built and analyzed to identify potential security and privacy threats. They do not, however, support a continuous discovery and threat analysis, since the DFD is built manually. 
%

Also, configuration monitoring tools and approaches, like GMonE~\cite{montes2013gmone} and Clouditor\footnote{\url{https://github.com/clouditor}}, exist. Yet, they lack the possibility to identify combinatorial CWs, since they just identify vulnerable configurations of single resources. 
A continuous threat assessment tool for multi-cloud systems is proposed by Torkura et al.~\cite{torkura2020continuous} who combine cloud monitoring with state transition analysis. Their approach, however, is also aimed at detecting threats based on expected configurations rather than comprehensive CWs. 


In summary, current literature does not sufficiently integrate regular expert-based threat analyses in the risk assessment process. Yet, we would argue that while expert analyses are costly, they are also essential in the assessment of security risks. One reason for this gap is that the analysis results need to be made reusable and automatically applicable across resources.
%
%
In the following sections, we address these gaps by making threat assessment results reusable and shareable in the form of policy-based threat profiles, and integrate them into a continuous process model.


\section{Process Model}
\label{process_model}

The cloud risk assessment methodology we propose in this section is targeted at organizations that build and operate cloud systems in public cloud infrastructures. 
Note that we assume a risk assessment methodology similar to how NIST describes it~\cite{blank2011guide}: A risk assessment consists of a threat assessment that identifies potential threat vectors for all assets, and evaluates their probabilities of occurrence. Then, impacts, e.g. financial and privacy impacts, of these attacks are evaluated. Finally, risks can be derived from the combination of threat probabilities and their potential impacts.

\textbf{Running example:}
Company \textit{Cloudara} sells smart home appliances, that are connected to Cloudara's cloud platform service. Security experts are worried about various risks on the infrastructure layer: cloud resources could be misconfigured allowing customers and outside attackers access to sensitive data, customers’ appliances could inject malicious data into the system, customer data may be lost due to missing backups, etc.
Moreover, there are several teams developing applications for the cloud platform who frequently change resource configurations. Additionally, subcontractors deliver code for cloud apps, and have temporary access to the platform, which is a hybrid system hosted using infrastructure provided by different infrastructure as a service (IaaS) providers. Cloudara's security experts have conducted a risk assessment and also manually monitor the platform, but are overwhelmed by the many changes: theoretically, any change needs to be evaluated in combination with all other configurations to check whether a change results in a new or modified risk.

\begin{figure}
\begin{center}
\includegraphics[width=0.9\linewidth, keepaspectratio]{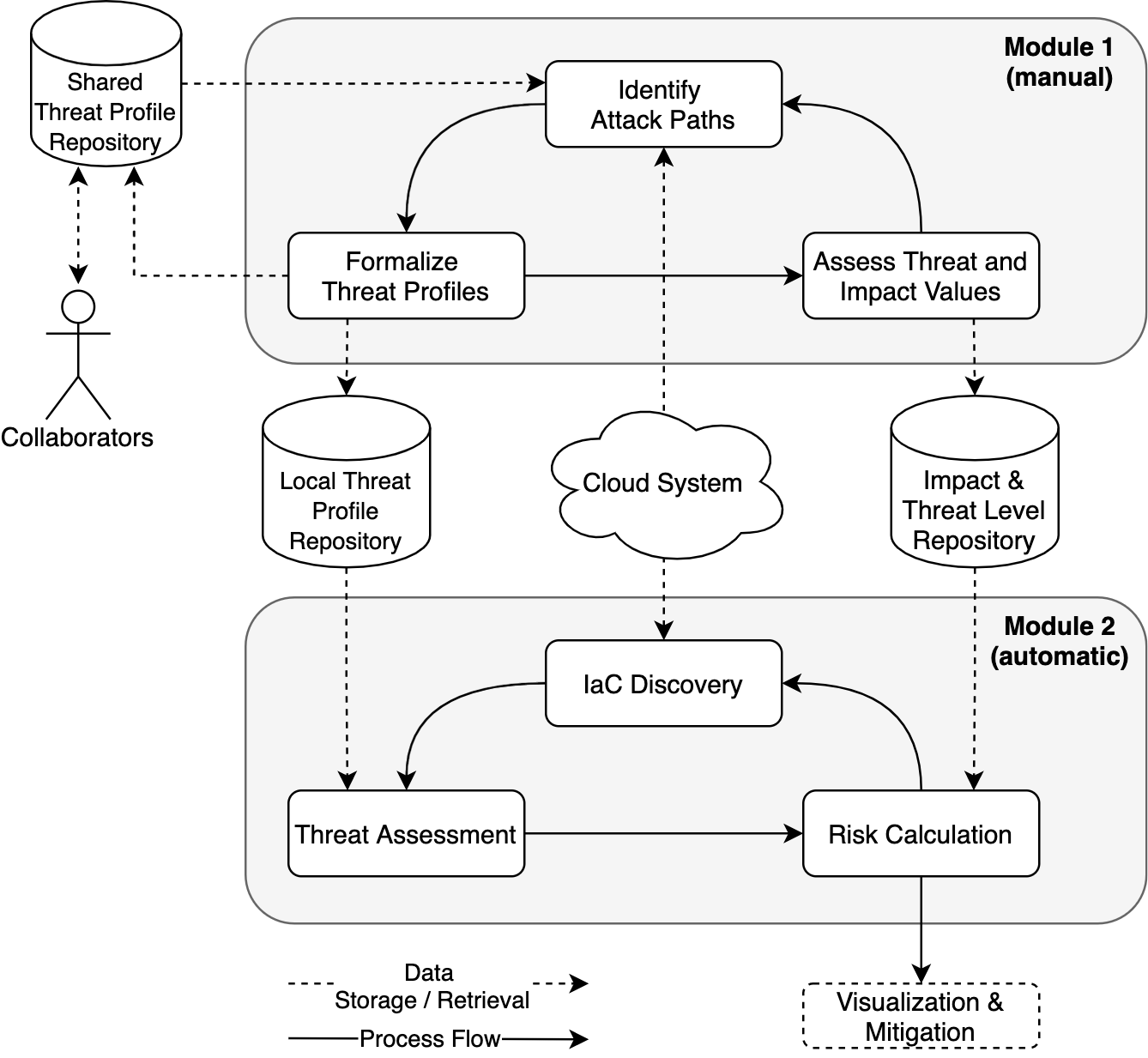}
\caption{An overview of the process model: We reorganize the traditional risk assessment phases in two modules of which the first one is conducted manually and the second one is completely automated. We also integrate collaboratively identified threats.} 
\label{fig:process}
\end{center}
\end{figure}

Figure~\ref{fig:process} shows an overview of the process model of our approach. It consists of two modules that together form the continuous risk assessment process.

\subsection{Module 1: Manual Preparation and Analysis}
\label{module1}
Module 1 covers some of the traditional steps to prepare a risk assessment, as well as a modified threat analysis that creates threat profiles. A \textit{threat profile} is the structured description of a number of infrastructure-level configuration weaknesses that enable an attack vector.  

Before the continuous process starts, the risk assessment activities are prepared. 
For instance, relevant assets, i.e. cloud resources that should be assessed, are identified. These can be any resources that can be provisioned and configured by a cloud user, like virtual machines, blob storages, and security groups. 
In AWS for example, an asset can be any resource of the AWS \textit{resource-types}\footnote{see \url{https://docs.aws.amazon.com/AWSCloudFormation/latest/UserGuide/aws-template-resource-type-ref.html} for a list of AWS resource types per service, and see \url{https://docs.microsoft.com/en-us/azure/azure-resource-manager/management/resource-providers-and-types} for guidance on how to achieve the same for Microsoft Azure.}. 

We take into account the three classic protection goals confidentiality, integrity, and availability for all assets rather than excluding one that is not relevant to an asset momentarily, since their analysis results are designed to be applicable for all future assets of that type also.

Also, the risk model is defined here. For example, scales for the quantification of threats and impacts, as well as risks, may be defined.
Since our methodology integrates shared threat descriptions, respective repositories are set up or identified which may be shared with the public, or with a selected set of organizations, e.g. across organizations. 

\subsubsection{Identify threat vectors}
In this activity, the traditional process of expert-based threat analysis is performed, i.e. threat vectors are identified and evaluated regarding their threat probabilities. 
Various methods to identify such threats are usable to that end. For example, attack trees can be created to identify potential attack vectors and derive weaknesses that would enable these attacks (see next section). In our approach we use attack trees as an example as explained in the following. 

\begin{figure}[t]
\begin{center}
\includegraphics[width=\linewidth, keepaspectratio]{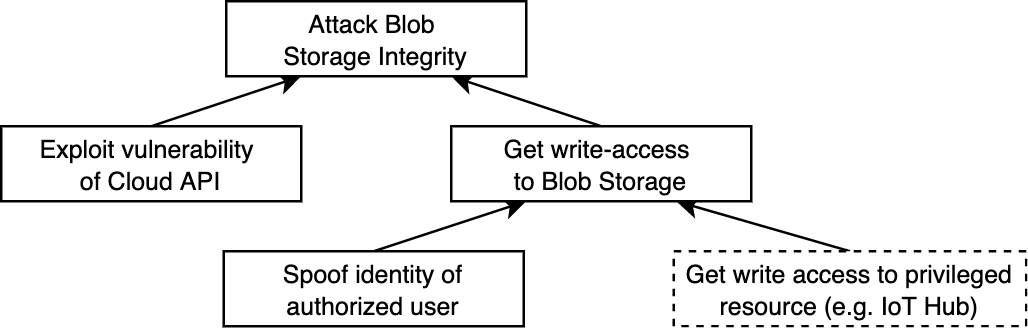}
\caption{An excerpt of the attack tree from our running example. The leaf node that refers to access via other privileged resources is a reference to respective attack tree of that respective resource. This way, complex attack vectors can be modeled.}
\label{fig:attacktree}
\end{center}
\end{figure}

\textbf{Running example:}
Cloudara's analysts create an attack tree for the integrity of a blob storage (see Figure~\ref{fig:attacktree}) which, e.g., reveals possible attacks via spoofing an authorized user's identity or injecting data via other privileged resources.


\subsubsection{Create threat profiles}
\label{sec:threatProfiles}
First, the new threat profiles from shared threat profile repositories are reviewed, adapted if necessary, and integrated into the local threat profile repository. 
%
When using attack trees, a CW can represent one leaf of an attack tree, or a combination of leaves that are connected via \textit{and} conditions. 

\begin{figure}[t]
\begin{center}
\includegraphics[width=0.7\linewidth, keepaspectratio]{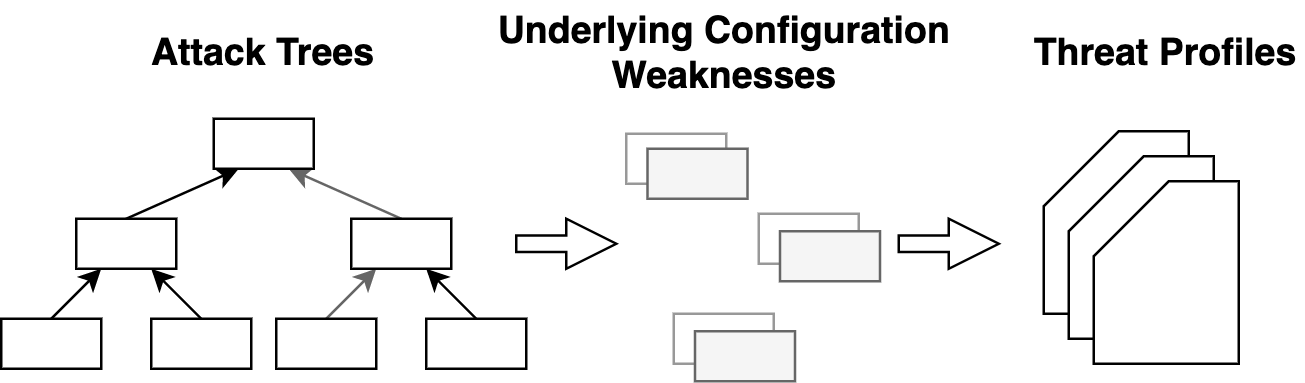}
\caption{Constructing machine-readable threat profiles: First, attack trees are constructed via an expert analysis. 
Every branch's underlying conditions, i.e. the CWs, are then described in a threat profile.}
\label{fig:threatprofiles}
\end{center}
\end{figure}

\textbf{Running example:}
The attack tree is now analyzed regarding its underlying weaknesses: For example, a bad password policy may enable the spoofing attack branch. In the following we focus on the branch that obtains access via another privileged resource: the company employs several IoT Hubs that accept traffic and store messages in the blob storage. The underlying configurations that enable this threat therefore include public reachability of the IoT Hub, and its access to the blob storage (see also Figure~\ref{fig:threatprofiles}). 
A corresponding threat profile is shown in Listing~\ref{lst:threatprofileexamples2}.

\begin{lstlisting}[
    basicstyle=\notsotiny,
    caption={A simplified example of a configuration weakness represented as a Rego policy: It checks whether any IoT Hub exists that has an access key ("connectionString") for a Storage Account configured---effectively checking whether a malicious IoT device could inject data into that Storage Account. Additionally, it checks whether the IoT Hub allows public access (last line).},
    label={lst:threatprofileexamples2},
    keywordstyle=\bfseries,
    captionpos=b,
    belowskip=0.2 \baselineskip
]
storageaccount_integrity_uploadFromMaliciousIoTDevice[account_names]{   
    input.resources[i].type == "Microsoft.Storage/storageAccounts"
    input.resources[j].type == "Microsoft.Devices/IotHubs"
    contains(input.resources[j].properties.routing.
                endpoints.storageContainers[_].connectionString,
             input.resources[i].name)
    input.resources[j].properties.publicNetworkAccess == Enabled}
\end{lstlisting}

Generally, our approach allows three different possibilities of defining threat profiles.
First, they can be written for specific cloud resources (or combinations thereof), e.g. configurations of a specific virtual machine and object storage. Example 1) in Listing~\ref{lst:threatprofileexamples}, for instance, shows a simple confidentiality threat description for a specific resource that results from an insecure endpoint configuration. 

Second, they can be written for cloud resource \textit{types} (or combinations thereof), e.g. configurations of virtual machines and object storages in general (example 2) in Listing~\ref{lst:threatprofileexamples}). 

Third, they can be written according to their classification in a cloud provider-independent ontology. 
In a previous work\footnote{We will add the reference here for the camera-ready version.} we have proposed such a cloud resource ontology which systemizes cloud resources and their security features, and defines relationships between them. For example, an Azure VM is classified as a \textit{Computing Resource}, which in turn inherits properties from \textit{Cloud Resource}. Similarly, an AWS S3 Bucket is classified as an \textit{ObjectStorage}, which in turn inherits properties from \textit{Storage}, and \textit{Cloud Resource} as well. The ontology then defines relationships from these concepts to security features, e.g. every \textit{Cloud Resource} offers a \textit{GeoLocation}, and every \textit{Storage} offers \textit{at-rest encryption}. 
This abstraction allows us to define threat profiles that are independent of cloud providers and cloud resource types. The example 3) in Listing~\ref{lst:threatprofileexamples} shows how general configuration options, like transport encryption, can be targeted independently of the resource type and cloud provider. 


\begin{lstlisting}[
    basicstyle=\notsotiny,
    caption={Simplified examples of threat profiles written in Rego, the policy language we use in our implementation (Section~\ref{prototype}). The policies each show a list of conditions that enable an eavesdrop attack. In Rego, a policy is evaluated to be true if all its conditions are true. In this example, it also returns the affected resources' names.},
    label={lst:threatprofileexamples},
    keywordstyle=\bfseries,
    % morekeywords={storageacc{}ount_confidentiality_eavesdropOnConnection, get_default_names},
    captionpos=b,
    belowskip=0.2 \baselineskip
]
1) resource_confidentiality_eavesdropOnConnection["vm1"]{   
    input.resources[i].name == "vm1"
    input.resources[i].properties.supportsHttpsTrafficOnly == false}

2) VM_confidentiality_eavesdropOnConnection[resource_names]{   
    resource_names := input.resources[i].name
    input.resources[i].type == "Microsoft.Compute/VirtualMachine"
    input.resources[i].properties.supportsHttpsTrafficOnly == false}

3) all_confidentiality_eavesdropOnConnection[resource_names]{   
    resource_names := input[i].name
    input[i].securityFeatures.transportEncryption != "enabled"}
\end{lstlisting}

Analysts have to choose the most appropriate option depending on the cloud system and identified threats: while threat vectors that are specific for certain resources are better targeted with the first option, and threat vectors that concern a single cloud system are well targeted with the second, threats for hybrid cloud systems are more adequately covered with ontological threat profiles.

It is possible that threat profiles overlap when they are defined using several of the presented options; they will not, however, conflict. Redundant definitions are also easily identifiable later in the assessment results.

Newly identified CWs may also be shared with collaborators by contributing them to the shared repository. To prevent the creation of duplicates in the shared repository, we propose following naming schema for threat profiles: 
\verb|<assetType>_<protectionGoal>_| \verb|<attackShortDescription>|. Further descriptions can be added in comments to exactly describe the CW. Note that the actual implementation of the CW does not need to be standardized.

\textbf{Running example:} Imagine furthermore that Cloudara is owned by a parent organization which holds several companies that develop cloud applications. Since their security experts are doing a lot of redundant work analysing similar threats, they benefit from knowledge sharing using a common repository of threat profiles. 

\subsubsection{Assess threat and impact values}
\label{sec:impacts}
In the final activity of Module 1, experts assess the threat values of the identified threat profiles and the potential impacts of a successful attack. For every threat profile, the probability of a successful attack is evaluated, e.g. a value of \textit{low} or a respective numerical value. Impacts are evaluated per asset and protection goal. For example, a qualitative impact value of \textit{medium} or a quantitative impact value of \textit{2.5} may be identified for a successful attack on the availability of a virtual machine. 
These assessments are stored in the Impact \& Threat Level Repository (see Figure~\ref{fig:process}).

Note that threat and impact levels can be omitted, if only potential threats shall be identified automatically. Missing threat and impact assessments are then reflected in the risk score (see below).

In summary, Module 1 consists of manually conducted activities that are repeated periodically. It is not the goal of this module to remove the necessity of expert analyses, but to design a risk assessment process that integrates them appropriately for assessing cloud systems, making their results reusable and automatically applicable.

There are different possibilities to define triggers for a repetition of Module 1. For example, a time-based trigger, like a monthly repetition, can be used. Also, new contributions to the shared threat profile repository may trigger a repetition since these may need to be reviewed and integrated. Also, a trigger from Module 2 may be used when the automatic assessment identifies resources that have no threat values or impact values defined yet.

\textbf{Running example:} The storage account is assessed with a medium impact level of $2$ since it holds valuable customer data of which, however, secure backups exist. 
To save time, Cloudara's experts furthermore apply this assessment to the storage's complete resource group since all resources in that group work with the data. The threat profile that describes the CWs for a data injection attack is also assessed with a medium value of $2$ because authentication mechanisms are in place, but customers' home networks may not be well secured. 

\subsection{Module 2: Continuous Application}
\label{module2}
Module 2 covers the continuous application of the previously defined threat profiles to the cloud system. 
We describe the activities of this module on a high-level here, and present an implementation in Section~\ref{prototype}.

\subsubsection{Resource discovery}
The first activity of Module 2 executes the discovery of existing cloud resources in the target system. The discovery results are forwarded to the next activity in a machine-readable description.

In this paper, we use Azure IaC descriptions, i.e. Azure Resource Manager (ARM) templates, for the discovery of cloud resources. Our approach, however, can also be used for other cloud systems, e.g. AWS and Google Cloud Platform (GCP). We discuss the respective differences to AWS and GCP in Section~\ref{comparison}. In Azure, IaC templates can be retrieved automatically via the respective APIs (see Section~\ref{prototype}).

In our implementation, we furthermore translate the IaC template to a cloud provider-agnostic representation that is based on the ontology of cloud resources described above (see also Section~\ref{prototype}). This way, we enable the assessment of threat profiles that target the ontology-based threat profiles.

\subsubsection{Threat assessment}
\begin{figure}[t]
\begin{center}
\includegraphics[width=0.6\linewidth, keepaspectratio]{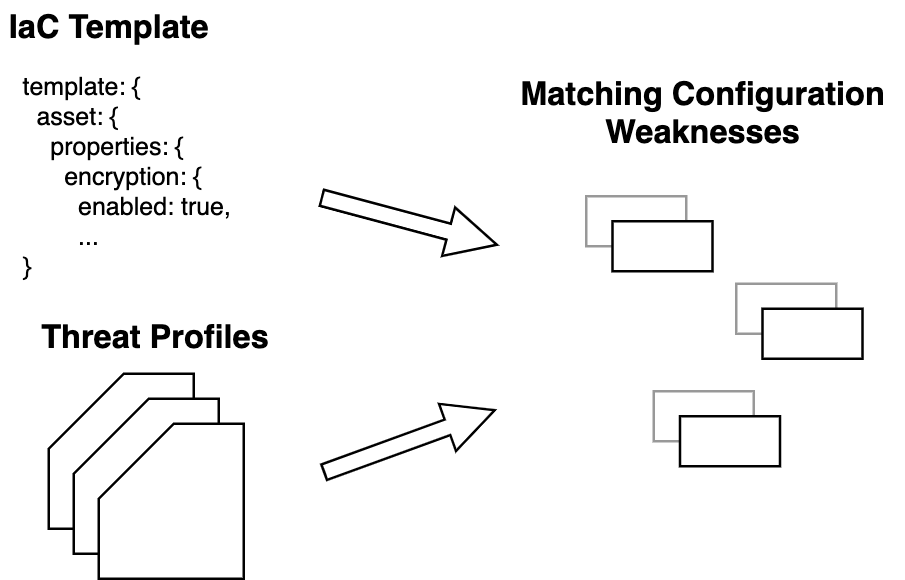}
\caption{Attack Tree Reconstruction: The current IaC template is evaluated against the threat profiles from the local threat profile repository and CWs for every resource are identified. Finally, these CWs are merged per asset to form a momentary attack tree.}
\label{fig:reconstruction}
\end{center}
\end{figure}

Next, the IaC templates are evaluated against the previously defined threat profiles, as depicted in Figure~\ref{fig:reconstruction}. For the task of this evaluation, a policy engine can be used. Usually, a policy engine evaluates incoming authorization requests against predefined policies and then approves or rejects the request. We use this mechanism for a different purpose, i.e. to evaluate cloud infrastructure descriptions (used as the authorization request) against predefined threat profiles (used as the predefined policies). 

The policy engine then outputs a mapping between CWs and a set of vulnerable assets from the input template. For instance, the output may indicate that a denial of service attack is possible for a list of resources, such as serverless functions or virtual machines, or that a data injection attack is possible for a storage container.
%
The outputs are reassembled construct a mapping of all applicable CWs per asset (see Figure~\ref{fig:reconstruction}) to make these results available for a possible manual inspection.

\subsubsection{Risk calculation}
Finally, risk scores are calculated using the previously defined impact and threat assessments stored in the Impact \& Threat Level Repository. 
Risk scores are then calculated according to the defined risk model. For example, numerical threat values and impact values may be multiplied to calculate the risk value per asset and protection goal. We do not, however, impose a specific way of calculating risks here, since our approach is independent from the underlying risk model.

In case no threat or impact values have been defined for a resource, the respective value is evaluated to $0$, allowing for a simple identification of resources that are lacking an evaluation. As mentioned above, this case can be used as a trigger for a new execution of Module 1.

In summary, Module 2 of the risk assessment process is completely automated and is performed independently of Module 1, creating an automatic assessment that can significantly improve the longevity of the manual activities. 
We envision Module 2 to be executed continuously, i.e. in short intervals, possibly of few minutes. Apart from this, it may be appropriate to trigger an execution based on changes in the cloud infrastructure, e.g. based on log data. The identification of such changes, however, is outside the scope of this paper.

\section{Prototype Implementation}
\label{prototype}

In this section, we describe our prototype implementation of the automated Module 2 which we instantiate for Microsoft Azure. It is implemented as an open-source project\footnote{\url{https://github.com/clouditor/continuous-risk-assessment}}. Considerations for extending the implementation to AWS and GCP are discussed in Section~\ref{comparison}. The prototype comprises the three activities \textit{discovery}, \textit{threat assessment}, and \textit{risk calculation}. Finally, we describe the results of our performance evaluation.
The prototype is implemented in Go and uses the policy engine \textit{Open Policy Agent} (OPA), its policy language \textit{Rego}\footnote{\url{https://www.openpolicyagent.org/docs/latest}} as well as parts of Clouditor\footnote{\url{https://github.com/clouditor/clouditor}} for cloud resource discovery. 

\subsection{Module 2 Implementation}
\subsubsection{Discovery}
In a first step, we discover existing cloud resources of the target cloud system as IaC templates, which in the case of Microsoft Azure are called ARM templates. The discovery step is specific to the cloud provider and in this example realized via APIs provided by Azure\footnote{\url{https://docs.microsoft.com/en-us/rest/api/resources/resourcegroups/exporttemplate}}, which allow to export ARM templates for single resources, complete resource groups as well as for a whole subscription. 
A sample excerpt of an ARM template is shown in Listing~\ref{lst:arm}.

\begin{lstlisting}[
    basicstyle=\notsotiny,
    label={lst:arm},
    caption={An excerpt of an ARM template. The \textit{resources} section includes all resources and their configurations.},
    keywordstyle=\bfseries,
    % morekeywords={$schema,contentVersion,parameters,variables,resources, properties},
    captionpos=b,
    belowskip=0.2 \baselineskip
]
{"template": {"$schema": "https://schema.management.azure.com/
        schemas/2015-01-01/deploymentTemplate.json#",
      ...},
      "resources": [{
        "type": "Microsoft.Storage/storageAccounts",
        "name": "risky_storage",
        "properties": {
            "azureFilesIdentityBasedAuthentication": {
                "directoryServiceOptions": "None"},
            "encryption": {
                "keySource": "Microsoft.Storage",
                "services": {"blob": {
                        "enabled": true,
                        "keyType": "Account"}}},
            "minimumTlsVersion": "TLS1_0",
            "supportsHttpsTrafficOnly": false, ...},...},...],...}
\end{lstlisting}

We then iterate through the resources defined in the template, and create a second resource definition according to the ontology definition, as shown in Listing~\ref{lst:ontologytemplate}.

\begin{lstlisting}[
    basicstyle=\notsotiny,
    label={lst:ontologytemplate},
    caption={An excerpt of the custom, ontology-based resource definition.},
    keywordstyle=\bfseries,
    % morekeywords={$schema,contentVersion,parameters,variables,resources, properties},
    captionpos=b,
    belowskip=0.2 \baselineskip
]
{"id": "/subscriptions/.../storageAccounts/risky_storage",
    "type": [
        "ObjectStorage",
        "Storage",
        "Resource"]
    "atRestEncryption": {
        "enabled": true,
        "keyManager": "Microsoft.Storage"},
    "httpEndpoint": {
        "transportEncryption": {
            "enabled": true,
            "enforced": false,
            "tlsVersion": "TLS1_0"}, ...}, ...}
\end{lstlisting}

\subsubsection{Threat Assessment}
In this step, the automatic threat assessment is conducted which primarily comprises the evaluation of the infrastructure description against the previously defined threat profiles.

For the evaluation of a given infrastructure description against predefined CWs, we use the previously discovered IaC template, and its ontology-based version, as the data input and the threat profiles defined in Module 1 as the policy input for the policy engine. 
The output is a mapping of threat profiles to vulnerable assets (in the following called \textit{identified threats}). 
Hereafter, the process is independent of the chosen cloud provider. The \textit{identified threats} are now used to output a mapping of CWs per asset. This output is both the input for the final risk calculation, and represents a valuable insight for manual inspection of potential weaknesses.

\subsubsection{Risk Calculation}
The risk calculation is performed according to the initially defined risk model. In our example we multiply the threat value of an asset's protection goal with the respective impact value---both of which are within the interval of $[1,3]$---resulting in a risk value in the interval of $[1,9]$. This calculation is done again using Rego policies. 

The inputs for the risk calculation policies are the \textit{identified threats}, as well as the threat and impact values defined in Module 1. The output then contains the risk scores for all assets and their protection goals which are calculated again using a Rego policy. 
This policy calculates the risk values for all assets and their protection goals according to the threat values that have been assessed before before.

\subsection{Performance Evaluation}

We conducted a benchmark test of the automatic threat assessment, i.e. the functionality that evaluates an IaC template against a set of policies, with varying sizes of templates and policies, since this is the part of the implementation that may be a bottleneck when the number of templates and policies increases.
The measurements were performed on an Azure VM Standard B2s with 2 vCPUs and 4 GiB memory. We used the Go Testing library for the measurement which also provides benchmarking functionality. Note that the tests are published along with our source code and threat profiles on GitHub.

\begin{figure}[t]
\begin{center}
\includegraphics[width=0.7\linewidth, keepaspectratio]{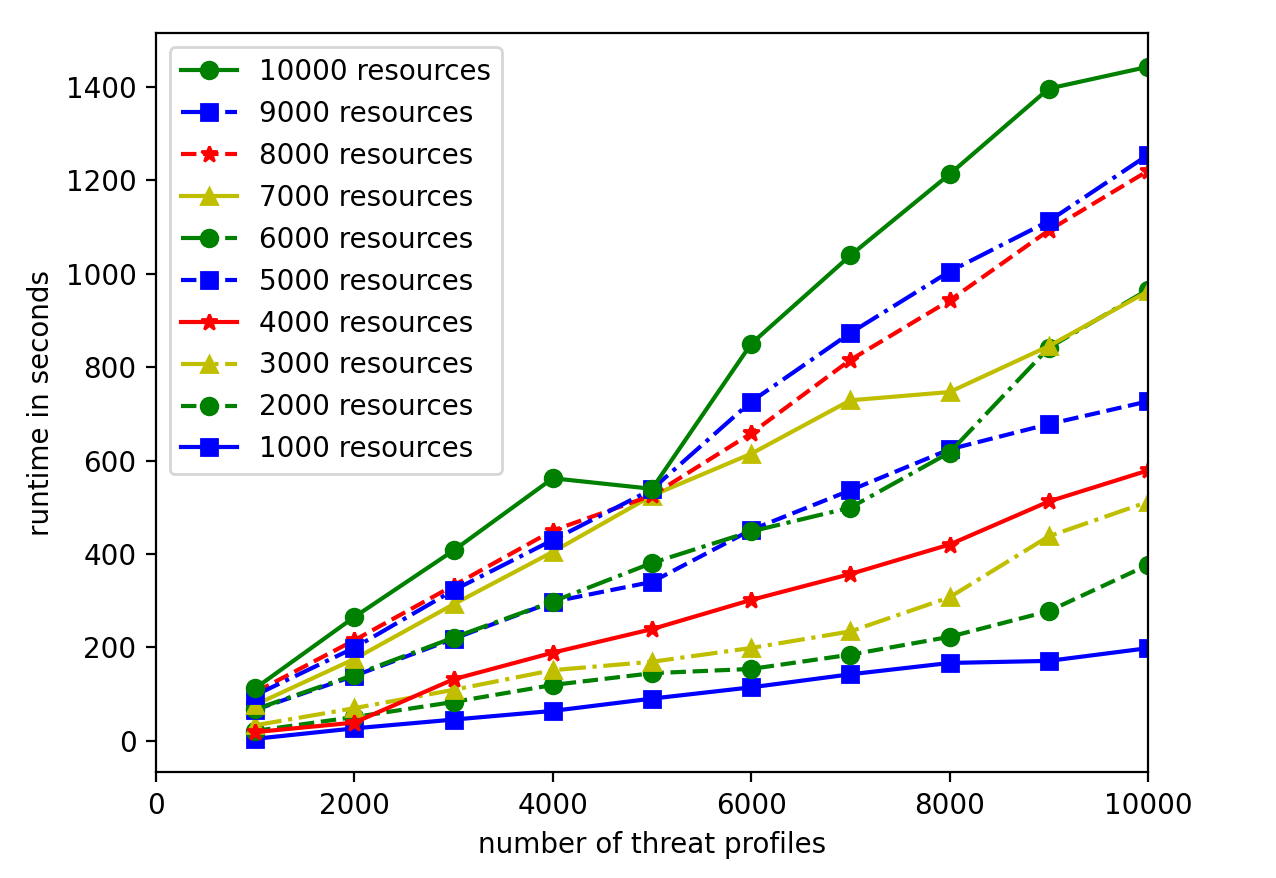}
\caption{Threat assessment performance}
\label{fig:threat_performance}
\end{center}
\end{figure}

We measured the performance of the threat assessment varying the number of resources of the IaC template between $1000$ and $10,000$. Likewise, we varied the size of the policies, or threat profiles, in the same range. Figure~\ref{fig:threat_performance} shows the measured runtimes.
Our benchmarking results show that the performance of the threat analysis increases linearly with the number of threat profiles and resources in the template. 
It can be seen that even in case both threat profiles and cloud resources need to be evaluated in high numbers, the application of our methodology using OPA is still feasible. 
%
The resources that were used in the benchmark were generic resources with a simple security property similar to firewall rules. The threat profiles that were used in the benchmark checked if this security property equals the String \textit{allow}.


Summarizing, we would argue that our approach allows to perform assessments of a cloud system's IaC description against a (large) number of threat profiles in frequent intervals of few minutes\footnote{For more details on the performance of OPA, see its documentation: \url{https://www.openpolicyagent.org/docs/latest/policy-performance}}.

\section{Discussion}
\label{discussion}
\label{sec:discussion}

\subsection{Module 1: Analysis and Threat Profiles}
\subsubsection{Manual analyses}
A limitation of our approach is that it does not completely alleviate the necessity for conducting manual risk assessments since the Module 1 activities still need to be performed periodically. It does, however, extend the longevity of manually performed analyses considerably by generating reusable threat profiles that can automatically be applied to newly created (or modified) resources. It is also an improvement towards an appropriate adaptation of risk assessment to cloud systems, proposing a systematic way to approach risk assessment in the cloud.
Furthermore, it integrates shared repositories that may include previously unidentified threat profiles as well as threat profiles for asset types that may be used in the future.

\subsubsection{Sharing and Reusing Threat Profiles}
The reusability of threat profiles across resources and organizations depends on how they are defined. As described in Section~\ref{sec:threatProfiles}, they can be defined for specific resources---limiting their reusability---or defined on a more abstract level---increasing their generality and reusability.
To improve reusability, we have defined a naming scheme in Section~\ref{module1}. Yet, the applied naming can still differ depending on the results of the manual threat analysis. 
Note, however, that in case of duplicate threat profiles, the final risk scores for the respective assets would not change as long as their probabilities are evaluated equally. 

Concerning the shared threat profile repositories, cloud users may see a risk in sharing their threat profiles with others, since they reveal information about the architecture of their system. Yet, contributing threat profiles to shared repositories also shows that a cloud user is aware of the respective CWs.
Also note that it would be possible to make anonymous contributions to a shared repository, e.g. using a form of mix networks, to obscure which participant has contributed which threat profiles.
The general problem of sharing security information across cloud users has been investigated in other works~\cite{kamhoua2015cyber,zhang2015community}.

\subsection{Module 2: Limitations of IaC Templates}
Our approach is furthermore limited by the properties that IaC templates offer, i.e. resource configurations on infrastructure-level. Risk assessments, however, may also include other types of CWs, such as social engineering or application-level threats. Note that such threats can still be integrated into the threat profiles, e.g. by adding a social engineering weakness to every resource of the system. 


Furthermore, when sharing threat profiles, in many cases, organizations will have to adapt the threat profile syntax to their own usage of IaC templates, since e.g. ARM templates can be customized with extra properties like variable names that may need to be integrated into such threat profiles. 
This approach is also limited by the fact that some resource types cannot be exported via the respective APIs\footnote{\url{https://docs.microsoft.com/en-us/azure/azure-resource-manager/templates/export-template-portal\#limitations}} (yet), e.g. the environment variables of Azure Function Apps.

In summary, Module 2 is limited by the possibilities that are provided by the cloud providers' IaC templates. We would expect, however, that cloud providers will expand the capabilities of such templates in the future as the usage of IaC is becoming more popular as part of the DevOps approach.

\subsection{Module 2: IaC Discovery in AWS and GCP}
\label{comparison}

The AWS CloudFormation service uses templates and so-called \textit{stacks} to manage IaC deployments. 
Exporting an existing AWS infrastructure as an IaC template is not as straightforward as in Azure. In AWS, IaC templates can be exported only if a template describing these resources has been created in another way beforehand (as discussed below)\footnote{\url{https://aws.amazon.com/blogs/aws/new-import-existing-resources-into-a-cloudformation-stack/}}.


In GCP, the Cloud Deployment Manager\footnote{\url{https://cloud.google.com/deployment-manager/}} can be used for describing resources for a deployment. 
In GCP, IaC templates are called \textit{configurations} and are written as YAML files, while the term \textit{templates} in GCP represents a reusable module of a configuration. 
Similar to AWS, the GCP Cloud Deployment Manager does not offer the possibility to export a template of existing resources or an existing resource group. 



In summary, the automatic discovery of resources in AWS and GCP differs significantly from Azure since they do not (yet) offer the possibility to automatically export an IaC template of existing resources. Using AWS or GCP with our methodology is, however, still possible if the templates are created in a different way. Consider that many organizations already define their infrastructures using IaC templates in which case an export from the existing cloud system is not necessary for the discovery. 
Popular tools that are used to manage the deployment of IaC templates automatically include Chef, Puppet, and Ansible\footnote{\url{https://www.chef.io/}, \url{https://puppet.com/}, \url{https://www.ansible.com/}}. 
Also, AWS and GCP may offer a similar export functionality in the future.

\subsection{Practical Feasibility}
The set up and maintenance of the Module 2 components can introduce an overhead in comparison to traditional security audits and infrastructure monitoring tools. 
The biggest overhead our approach introduces may be the maintenance of threat profiles which can be necessary for several reasons.

First, there may be changes in the configuration options provided by the cloud provider which should be reflected in existing threat profiles. In comparison to other security monitoring alternatives, however, our approach is not more or less laborious, since service providers always need to keep up with the changes a cloud provider introduces. 
Second, threat profiles may be deprecated due to other reasons, e.g. when a certain configuration is not considered vulnerable anymore. Consider, however, that such profiles can be changed or deleted once they alert a risk that is considered a false-positive, rather than maintaining all threat profiles periodically. 
Third, sharing threat profiles may be tedious if the respective repositories become large. Therefore we have discussed possibilities to further standardize the definition of threat profiles (see above), and we allow for the definition of generic, ontology-based threat profiles.
Note also that a similar idea to our proposal of a shared repository of cloud vulnerabilities has recently been discussed in the cloud security community. 

\subsection{Effectiveness}
Table~\ref{tab:comparison} gives a comparative overview.

In comparison to other threat and risk assessment approaches, we would argue that our approach combines the advantages of manual risk assessments with automated monitoring solutions. Consider the following examples:

\subsubsection{Temporary resources}
In cloud systems it may, for instance, be practical to implement temporary jump servers to enable easy access to application resources, e.g. for external partners. In this case a VM with SSH access may be set up and combined with a network security group that makes port 22 publicly available. Temporary, vulnerable resources like these would quickly be identified in our approach---independently of its resource type.

\subsubsection{Conditional threat vectors} 
Often, automatic threat identification results in many false positives, since they depend on certain resource properties. For example, cloud service providers often employ different environments with varying security requirements, e.g. different resource groups for development and production.
In the threat profiles we have proposed, policies can easily be scoped to certain environments, like resource groups, or single resources.

A further example is the location management of resources. While usually, resources and sensitive data should be kept in the same geographic region, e.g. due to the restrictions of data protection regulations, there are cases in which a cloud service provider may move data, like backups, to another geographic location. In our approach, a specific threat profile can be defined that identifies risky storage resources by evaluating their location in combination with their encryption mechanism: if it uses customer-managed keys it may be considered a low threat even if it is in a disallowed location.



\subsubsection{Cross-resource threats} 
We can also represent threat vectors across many resources, possibly independently of their resource type, which is not possible in standard cloud monitoring, and may be difficult even in manually conducted assessments. For example, we can identify a threat to any storage resource that results from any computing resource that is connected to a networking resource which in turn opens up a vulnerable port---combining many resource properties.



\subsubsection{Hybrid clouds}
Since we also allow to define threats based on all levels of a cloud resource ontology, we simplify the security management of hybrid cloud systems. Ontology-based threat profiles are not only applicable to several cloud providers, but can also represent threat vectors across cloud providers, e.g. if a threat vector is identified across a computing resource in one cloud system, which has access to a storage resource in another.


In summary, we expect our approach to be significantly more effective than traditional assessments since it makes expert analysis results applicable continuously across resources. Thus, it allows for a quick identification of risks as described above, also if they only occur temporarily. 
Also, it results in less false positive alerts than standard infrastructure monitoring tools which usually only monitor single configurations or single resources rather than combinations of configurations. 
%
Our approach would also recognize these risks unambiguously due to the standardized format of IaC and ontology-based resource descriptions which presents another advantage over automated monitoring solutions.


\newcommand{\pie}[1]{%
\begin{tikzpicture}
 \draw (0,0) circle (1ex);\fill[rotate=90](1ex,0) arc (0:#1:1ex) -- (0,0) -- cycle;
\end{tikzpicture}%
}

\rowcolors{2}{gray!25}{white}
\begin{table*}
  \caption{Comparison of cloud monitoring tools: DARGOS, Sensu, and GMonE are taken from Ward and Barker, the others have been added by us. We have furthermore added the criterion of resource-/vendor independence.}
  \centering
  \label{tab:comparison}
  \begin{tabular}{>{\centering\arraybackslash}m{0.55in} | >{\centering\arraybackslash}m{0.55in} | >{\centering\arraybackslash}m{0.55in} | >{\centering\arraybackslash}m{0.55in} | >{\centering\arraybackslash}m{0.55in} | >{\centering\arraybackslash}m{0.55in} | >{\centering\arraybackslash}m{0.55in} | >{\centering\arraybackslash}m{0.55in} | >{\centering\arraybackslash}m{0.55in}}
  \rowcolor{gray!25}{}
   \textbf{System} & \textbf{Scalable} & \textbf{Cloud-Aware} & \textbf{Fault-Tolerant} & \textbf{Multiple-Granular} & \textbf{Compre-hensive} & \textbf{Time-Sensitive} & \textbf{Autonomic} & \textbf{Resource-/Vendor-Independent} 
   \\
   \hline
   DARGOS	& \pie{360} & \pie{360} & \pie{0} 	& \pie{360} & \pie{0} 	& \pie{360} & \pie{0} 	& \pie{0} 	\\ 
   Sensu	& \pie{360} & \pie{0} 	& \pie{0} 	& \pie{0} 	& \pie{0} 	& \pie{0} 	& \pie{360} & \pie{0} 	\\
   GMonE	& \pie{360} & \pie{360} & \pie{0} 	& \pie{0} 	& \pie{360} & \pie{0} 	& \pie{0} 	& \pie{0} 	\\
   MRA	   	& \pie{0} 	& \pie{360} & \pie{360} & \pie{360} & \pie{360} & \pie{360} & \pie{0}	& \pie{360} \\
   CRA	   	& \pie{360} 
   						& \pie{180} 
   									& \pie{360} 
   												& \pie{0} 
   															& \pie{180} 
   																		& \pie{0} 
   																					& \pie{360} 
   																							 	& \pie{360} \\
  \end{tabular}
\end{table*}

While our approach is a risk assessment methodology, its main contribution lies in the automated module that functions similarly to cloud monitoring tools. We therefore present Table~\ref{tab:comparison} which adds our approach (CRA), as well as manual risk assessment (MRA), to the comparison done by Ward and Barker~\cite{ward2014observing}. In their work they compare different cloud monitoring tools according to the following criteria. 

\textbf{Scalability} refers to the ability of a system to ``adapt to elasticity''. As shown in the performance evaluation, the automated module of our approach scales well with the increasing size of resources.
\textbf{Cloud-awareness} refers to a system that minimizes delays and costs, e.g. by being location-aware of VMs. Our approach is not cloud-aware in this sense, since it is not ``aware'' of resource locations. We would argue, however, that this criterion is not relevant for our approach, since the data traffic our approach generates does not incur costs---the relevant API calls are free of charge---and because the data collection is done by the cloud provider---which we assume could not be done more efficiently with a self-deployed system.
Our approach is furthermore \textbf{fault-tolerant}, since it discovers available resources rather than expecting ones which may not be available.
\textbf{Multiple granularities} are given if a system presents visualizations on different levels of detail, and if it taps different sources for data collection, e.g. different layers of the cloud stack. Our approach does not fulfill this criterion (yet); we regard the addition of visualizations and the addition of threat detection on, e.g., the software layer as future work.
Our approach is also \textbf{comprehensive} in some respects, as it covers different cloud providers and services. It does not, however, use other sources for monitoring data, such as hardware and operating systems, which we regard as out of scope for this paper.
\textbf{Time-sensitivity} refers to the property of providing time guarantees or reducing latencies. Again, this is not covered by our approach, but we would argue that it is not relevant since the resource discovery is done by the cloud provider. Similarly, we expect latencies to be low as long as the cloud provider conducts the data collection.
\textbf{Autonomic} systems ``require no significant configuration or manipulation at runtime'' which is fulfilled by our approach, since manipulations and configurations are completely decoupled from the automated monitoring and assessment steps.
With our approach we have introduced two new criteria, namely \textbf{resource-/vendor-independent} assessments, i.e. expected states can be defined independently of the resource type and cloud provider.



\section{Conclusions}
\label{conclusions}

Mitigating security problems quickly is especially important in the cloud where changes on the infrastructure-level occur frequently. The automation of risk assessment allows to keep up with these changes.

We have proposed a redesigned risk assessment methodology that partly automates the risk assessment process for cloud systems. It employs reusable threat profiles that can be applied across assets, and it integrates shared threat profile repositories that allow organizations to collaborate on threat profiles. Such a collaboration can speed up the threat analysis of new resources and configurations, and can complement existing ones.
We have supported our methodology by an extendible prototype implementation that continuously applies threat profiles to cloud resources using the Open Policy Agent. 
Our implementation is built in a modular fashion allowing for adaptations to other requirements, e.g. changes to the risk model.
We envision our methodology being used in a collaborative way by the cloud community, as well as between organizations. To that end, we have created a public shared repository that can be used and contributed to by others. 

In future work, we want to bridge the gap between infrastructure-level and application-level risk assessment by integrating static code analysis methods in our approach.
We also plan to extend our methodology and threat profile repository, e.g., with Kubernetes resources, and add suggestions for mitigations. 



\bibliographystyle{unsrt}
\bibliography{bib}

\end{document}